\documentclass[letterpaper, 10 pt, conference]{ieeeconf}
\usepackage{CJKutf8}
\usepackage{comment}
\usepackage{amsmath}
\usepackage{tabularx}
\usepackage{booktabs}
\usepackage{multirow}
\usepackage{mathtools}
\usepackage{algorithm2e}
\usepackage{graphicx}
\usepackage{amssymb, mathtools}
\usepackage{capt-of}
\usepackage[pass]{geometry}
\usepackage{xcolor}
\usepackage{balance}
\usepackage{CJKutf8}
\usepackage{url}
\input epsf
\usepackage{graphicx}
\usepackage{colortbl}
\usepackage{algorithm}
\usepackage{algorithmicx}
\usepackage{algpseudocode}
\usepackage{xcolor}

\usepackage[T1]{fontenc}
\usepackage[scaled=0.85]{beramono}
\usepackage{listings}
\lstdefinestyle{SQL}{
    language=SQL,
    basicstyle=\ttfamily,
    keywordstyle=\color{teal},
    commentstyle=\color{gray},
    stringstyle=\color{orange},
    morekeywords={SELECT, FROM, WHERE, GROUP BY, ORDER BY, INSERT INTO, VALUES, CREATE TABLE, VARCHAR, INT, PRIMARY KEY, FILTER}
}

\begin{document}


\title{Time-Probability Dependent Knowledge Extraction in IoT-enabled Smart Building}

\author{\authorblockN{Hangli Ge, Hirotsugu Seike, Noboru Koshizuka}
\authorblockA{Interfaculty Initiative in Information Studies\\
The University of Tokyo,
Tokyo, Japan \\
Email:  \{hangli.ge, hirotsugu.seike, noboru\}@koshizuka-lab.org
}}

\maketitle

\begin{abstract}
Smart buildings incorporate various emerging Internet of Things (IoT) applications for comprehensive management of energy efficiency, human comfort, automation and security. However, the development of a knowledge extraction framework for human activities is fundamental. Currently, there is a lack of a unified and practical framework for modeling heterogeneous sensor data within buildings. In this paper, we propose a practical inference framework for extracting status-to-event knowledge within smart building. Our proposal includes IoT-based API integration, ontology model design, and time probability dependent knowledge extraction methods. We leveraged the Building Topology Ontology (BOT) to construct spatial relations among sensors and spaces within the building. Additionally, we utilized Apache Jena Fuseki's SPARQL server for storing and querying RDF triple data. Two types of knowledge could be extracted: timestamp-based probability for abnormal event detection and time interval-based probability for conjunction of multiple events. We conducted experiments over a 78-day period in a real smart building environment, collecting data on light and elevator states for evaluation. The evaluation revealed several inferred events, such as room occupancy, elevator trajectory tracking, and the conjunction of both events. The numerical values of detected event counts and probability demonstrate the potential for automatic control in the smart building. 
\end{abstract}

\begin{keywords}
Smart Building; Knowledge Extraction; Ontology; Internet of Thing (IoT); Time Probability;
\end{keywords}

 
\section{Introduction}
The emergence of Internet of Things (IoT) technology \cite{atzori2010internet} has revolutionized the capabilities of buildings by integrating diverse cybernetic systems. These systems facilitate real-time monitoring and control of various building components, including lighting, HVAC (heating, ventilating, and air conditioning), alarm systems, surveillance cameras, power meters, and more. Additionally, the incorporation of artificial intelligence (AI) offers a myriad of innovative applications, such as supervisory automation, occupancy comfort optimization, energy efficiency, and security management. This integration effectively transforms traditional buildings into smart environments, where human and cybernetic systems synergize to create a cohesive and adaptive ecosystem.

However, implementing automatic knowledge extraction in smart buildings requires significant manual effort and domain expertise specific to the building. Despite the recent advancements made by large language models (LLMs) \cite{zhao2023survey} in comprehending complex data, a key challenge for smart environment persists due to the absence of data models of the physical world. Fundamentally, integrating heterogeneous devices and modeling the context, which encompasses users, sensors, authorities, locations, events, etc., alongside the building structure, is a particularly challenging task. The most significant obstacle is ensuring compatibility among various sensors and devices, each of which may offer different networking features, protocols, and interfaces from various vendors. Therefore, information transition among these heterogeneous sensors is costly and hampers the swift implementation of smart buildings. 

Furthermore, it hinders the representation of resources in a machine-readable format, lacks the ability to programmatically explore different aspects of a building. It also obstructs knowledge extraction with a human-readable description. For instance, when deploying a tracking application for security management, inquiries may arise such as ``\textit{Who was the first person to enter the building yesterday? At what time did he/she arrive? and which room did he/she enter}". This process involves the following information: 1) Identifying sensors (e.g., smart locks or surveillance camera footage) assigned to building's entrance; 2) Querying the first entry event yesterday and retrieving the timestamp; 3) Investigating the elevator movements within the short time frame surrounding the timestamp to determine the destination floor; 4) Querying the rooms on that floor to identify any status changes of the specific room. 

To achieve the above-mentioned knowledge extraction goals, the design of a unified API for interoperability among IoT devices becomes a crucial aspect. Additionally, integration by an ontology-based model is also essential for establishing spatial and contextual relations among sensors, spaces, and users, thereby facilitating automatic event extraction. Subsequently, knowledge extraction regarding event conjunction, considering continuous changes among events is able to be explored. In this paper, we introduce our proposal, a practical framework for event extraction in IoT-enabled smart building. Our contributions include:
\begin{itemize}
\item To address the interoperability challenge among IoT devices, we developed a unified IoT-open API, bridging the gap among various IoT sensory hardwares. 
\item Additionally, we employed the Building Topology Ontology (BOT) to model the spatial hierarchy of sensor and device nodes within the building. 
\item We proposed time probability dependent knowledge extraction methods for extracting anomaly detection and event conjunction detection. 
\item We conducted the experiment in a real smart building and collected numerical values of detected event counts and probability values, demonstrating the potential for automatic control in the smart building
\end{itemize}

\section{Related Work}
\subsection{Modeling Tools of Smart Building}
Ontology plays a vital role in specifying domain-specific concepts among the sensors and other building assets. Numerous research projects of semantic models aimed at facilitating building management. Most of these ontologies focus on realizing particular applications such as energy management \cite{degha2019intelligent,petrushevski2017semantic}, or automated design and operation \cite{ploennigs2012basont,rasmussen2019bot, ge2021applying}. These projects include rule-based supervision methods \cite{tamani2018rule}, definition of a metadata schema for building management \cite{gao2015data,balaji2018brick}, and the establishment of an ontology to incorporate ambient intelligence in smart buildings \cite{stavropoulos2012bonsai}. An ontology utilizing vocabulary-based approach, defines concepts and relationships through RDF (Resource Description Framework) \cite{lassila1998resource}. Several specific ontologies also have been proposed for the domain of smart homes and buildings. Notable among these are the Sensor, Observation, Sample, and Actuator (SOSA) ontology \cite{janowicz2019sosa} and the  Semantic Sensor Network (SSN) ontolog, both recognized as W3C recommendations, which provide a framework to describe hardware, observation of physical entities and actuation and more. The Building Topology Ontology (BOT) \cite{rasmussen2019bot} delineates relationships between the sub-components of a building. It is also adhering to general W3C principles, and is suggested as an extensible baseline for reuse.

\subsection{Detection and Reasoning in Smart Building}
Various models have been employed to detect the key variables influencing system dynamics. Most existing solutions use machine learning (ML) for smart building applications and it mainly focused on the occupant, including occupancy detection \cite{djenouri2019machine}, estimating users' preferences and identification \cite{djenouri2019machine}, and activity recognition \cite{ahmadi2017one}. For instance, Khan et al \cite{khan2014occupancy} dealt with occupancy of premise range from binary occupancy (occupied or out of occupied), categorical and exact numbers. The method integrated several types of sensors, including acoustic noise, PIR (Passive Infrared Ray), humidity, light and so on.  Similarly, existing studies \cite{kitagishi2022wi,choi2020iot,ge2022accurate} focusing on localization, human activity recognition in a small space with limited devices or sensors have also been proposed. 

However, these proposals require vast amounts of labeled data to ensure the supervised learning approaches are effective, which is not always possible. Moreover, most of these ML-based works focused on the detection within limited spaces, where the physical structure of sensor deployment was ignored. Therefore, to reduce the complexity of information across different domains and aim for the knowledge extraction in smart building, there are significant technical challenges remains to be addressed: 1). Integration of sensor or device nodes, which are decentralized in both cyber and physical dimensions, varying with its parameters; 2). Development of a semantic approach, utilizing relevant description logics (DL); 3). Developing mathematical proposal for event extraction that considers continuous changes across the entire environment.

\section{Proposal}

We present a framework for extracting knowledge for IoT-enabled smart buildings, which takes into account time probability dependencies. Our proposal consists of three main components: 1) a unified IoT sensor network for collaborative computing, 2) an ontology model for representing the semantic and physical relationships, and 3) time probability dependent knowledge extraction methods.


\subsection{Unified API design}
The unified IoT API encompasses the following  properties: (1) ucode \cite{koshizuka2010ubiquitous}: used as the unique identifier of the object; (2) name: assigned with th e name or description of the node; (3) data: composed of the sub-properties of `instance' and `time', in which `instance' represents the sensed value and `time' indicates the timestamp. RESTful GET method was utilized for receiving monitored status information utilizes, with the `ucode’ is required to be given whereas the time duration is optional.

\subsection{Ontology model and Sparql query language}
\begin{figure}[h]
\centering
\includegraphics[height=2in, width=1\linewidth]{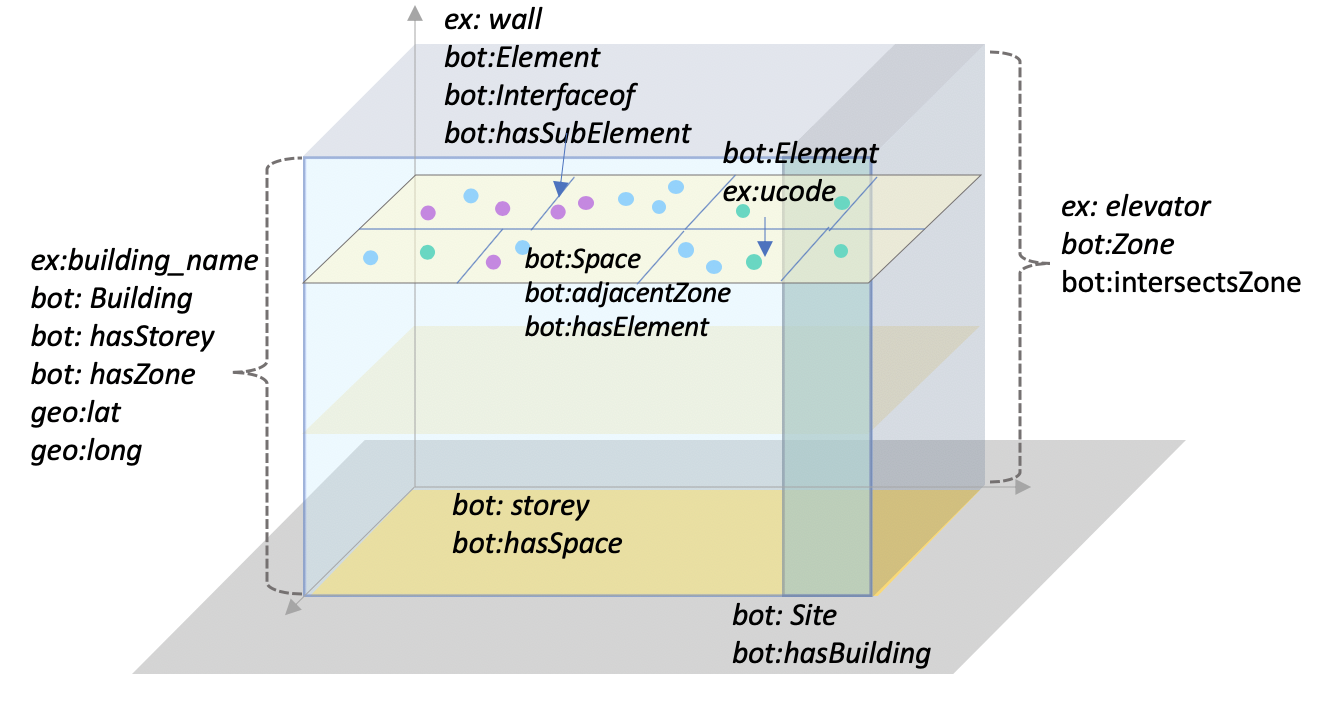}
\caption{The ontology structure of BOT}
\label{fig:kg}
\end{figure}



To align with and fully leverage established standards,
we incorporate the Building Topology Ontology (BOT). It was implemented based on OWL DL (description logic) ontology \cite{world2012owl}, and serves to define both semantic and physical relationships among the sub-components of a building. As illustrated in Fig~\ref{fig:kg}, we have designated Floor, Room, Zone, Sensor, Device, etc. as classes within the Building Topology Ontology (BOT). This semantic information is hierarchically structured to describes relationships, such as the type determinations of sensor associated with a room, the variety of spaces on a floor and the floors connected by elevator.


\begin{itemize}
\item Site: An area contains buildings.
\item Building: The designated building class allows referencing the entire context within it.
\item Storey: A level or floor part of a building. 
\item Space: A segment of a storey with a 3D spatial extent, bounded physically or theoretically. It serves specific functions within the storey, representing common spaces like rooms, corridors, elevator halls, toilets, etc. The space node assembles all sensor/actuator nodes within it.
\item Element: Sensor, actuator or device could be  defined as an element node which is deployed in a space. 
\end{itemize}

\begin{lstlisting}[style=SQL,captionpos=b, caption=Example1 of SPARQL query, label=lst:sparql,
basicstyle=\ttfamily,frame=single]
SELECT ?level ?space ?ele ?ucode   
  WHERE {    {   
    daiwa_bot:Elevator bot:intersectsZone ?level.   
    daiwa_bot:Elevator daiwa_bot:ucode ?ele_ucode.   
    ?level bot:hasSpace ?space.   
    ?space bot:hasElement ?ele.  
    ?ele daiwa_bot:element_type "Light".  
    ?ele daiwa_bot:ucode ?ucode .   
    FILTER (?level = daiwa_bot:Level1) 
  }  }   
\end{lstlisting}

For representing and querying ontology graphs, the SPARQL query language \cite{seaborne2008sparql} was employed, and we chose Apache Jena \cite{apache} as the platform for storing ontology data using the RDF data structure. Apache Jena, an open-source framework, facilitates the management and querying of RDF data. It incorporates a web frontend (Fuseki) and a SPARQL backend of TBD, which is a component of Jena for RDF storage and query. For instance, Listing.~\ref{lst:sparql} presents a query designed to retrieve `light' elements within rooms on the floor connected to `Level1' via the elevator. 

\subsection{Knowledge Extraction}
The knowledge base contains all the assertions and schema information pertinent to the domain, including dynamic data from the sensor nodes. The continuous real-time submission of data by nodes results in a large and frequently updated data store. Therefore, we have segregated the static ontology and dynamic sensing data into RDF store and relational database store, respectively. This approach aims for efficient data processing and high-level semantic integration.


\subsubsection{\textbf{Timestamp-based Probability for Abnormal Event Detection}}
Sensor or device states change due to user activities and their moving trajectories. It enables the probabilistic inference of human indoor mobility events through sensor observations. In this research, we constructed the knowledge graph of $G = (E, R, S)$, where $E=\{e_{1},e_{2},e_{3},...e_{n}\}$, represent the set of sensor nodes or devices, and $R$ denotes relations among elements. The corresponding data triple of ($e_{i}$, $S_{i}$) comprise the sensor ($e_{i}$) and its data observation ($S_{i}$) retrieved from the unified IoT-enabled API. 

Leveraging the aforementioned ontology of smart buildings and the continuously evolving sensor data enables knowledge extraction of various events. For example, consider sensor observations $S$ on the light of a room, where $S$ = $\{(s_{1},t_{1}), (s_{2},t_{2})...(s_{x},t_{x}) \}$, $s_{x} \in \{0,1\}$ denotes the value of room usage state. Subsequently, a  probability density function is utilized to describe semantics. For instance, query of `Is the room occupied abnormally?' can be deduced from the `light\_on' event considering the timestamp ($t_{x}$) when the event occurred. The probability of $t_{x}$ falls within any interval of $\left [ T_{1}, T_{2} \right ]$ (where $T_{2}$ - $T_{1}$ = 1 (hour)). Instead of considering individual timestamps, they 
are down-sampled to hourly units. Therefore timestamps are sampled into 24 value units. Variable of $P({e,t})$ represents the probability of the event happening at $t_{x}$, denoted by Equation~\ref{eq:p}:
\begin{equation}
P(T_{1}\leq t_{x} \leq T_{2}) =\int_{T_{1}}^{T_{2}} f(t_{x}) dt
\label{eq:p}
\end{equation} where $f(x)$ is the probability density function of the event.


Based on timestamp probability,
a simple rule for knowledge extraction of an event being normal or abnormal is denoted in Equation \ref{eqn2} :
\begin{equation}
\label{eqn2}
    S(event_{x},t_{x})= 
\begin{dcases*}
    \mbox{ \textit{Abnomal}}  & $P(event_{x},t_{x}) < \bar{p}$ \\ 
    \mbox{ \textit{Nomal}}   &otherwise.
\end{dcases*}
\end{equation} 
where $P(event_{x},t_{x})$ denotes the probability of $event_{x}$ occurring at $t_{x}$, $S(event_{x}, t_{x})$ denotes status of the event as normal or abnormal, and $\bar{p}$ is a threshold value that can be calculated based on historical data.

\subsubsection{\textbf{Time Interval-based Probability for Multiple Event Conjunction}}

Building on the single event detection, we have also proposed a time interval-based probabilistic model for detecting conjunctions of multiple correlated events. It creates high-level semantics. For example, when a user enters a building and takes an elevator to the destination floor, and then head to the target room, they activate multiple sensors and devices along the path. In particular, as illustrated in Figure~\ref{fig:sequece}, when a user intends to use the elevator:
\begin{figure}[h]
\centering
\includegraphics[height=1.1in, width=1\linewidth]{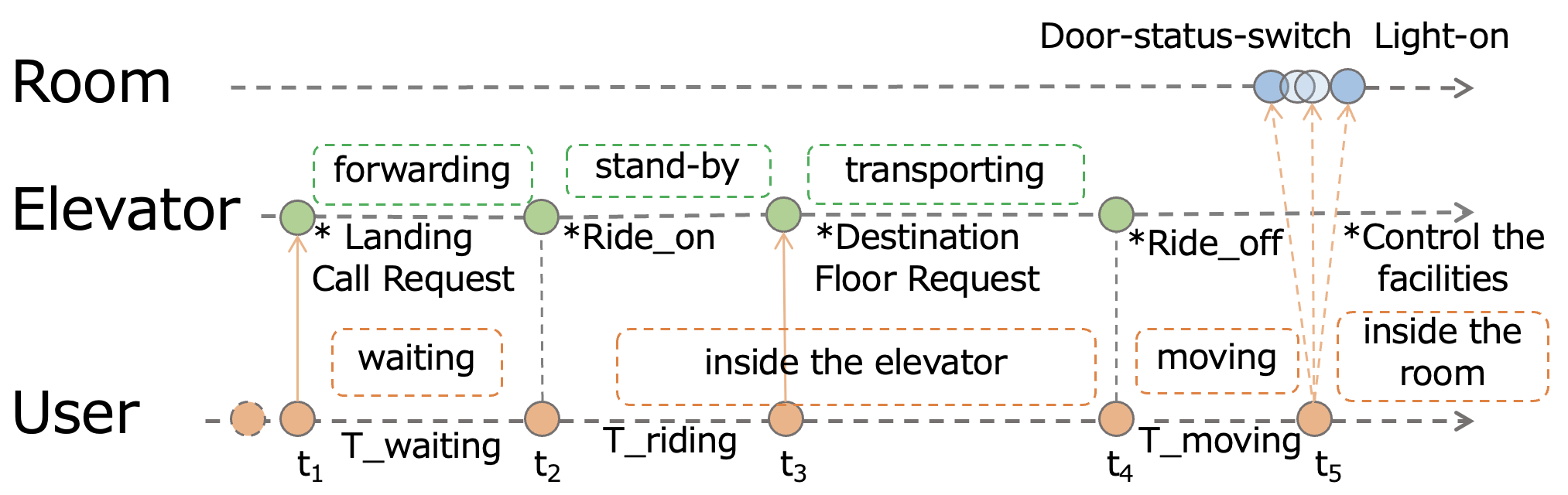}
\caption{Scenario of status transitions of room, elevator and user on riding on-off elevator and walking to the room}
\label{fig:sequece}
\end{figure}
\begin{itemize}
\item  The user initiates a landing-call request at $t_{1}$, and the elevator moves to the floor where the user is located. Meanwhile, the user continues to wait from $t_{1}$ to $t_{2}$.
\item The user enters the elevator at $t_{2}$, makes a destination floor request at $t_{3}$.
\item The user exits the elevator at $t_{4}$ after it arrives at the destination floor.
\item As the user proceeds to their room, entering the room may involve controlling facilities and triggering various sensor data around $t_{5}$ time.
\end{itemize}


This scenario accounts for various human activity contexts, such as entering or existing an elevator or turning on the light. By incorporating these contextual factors, the proposed model aims to provide a more holistic understanding of complex scenarios. Therefore, let $P(\Delta t_{x})$ denotes the probability of the estimated occurrence of event i ($E_{i}$) and event j ($E_{j}$)  with the given time interval $\Delta t_{x}$, where $\Delta t_{x} =t_{i}-t_{j}$. As denoted by Equation~\ref{eqn3}, given the probability distribution $P(\Delta t)$, $S(E_{i},E_{j},\Delta t_{x})$ represents the relation status of the two events of $E_{i}$ and $E_{j}$, and $\bar{p}$ is a threshold value calculated from historical data.\begin{equation}
\label{eqn3}
    S(E_{i},E_{j},\Delta t_{x})= 
\begin{dcases*}
    \mbox{ $E_{i}$ is \textit{Conjunctive to} $E_{j}$}  & $P(\Delta t_{x})  >= \widetilde{p}$ \\ 
    \mbox{ $E_{i}$ is \textit{NOT related to} $E_{j}$}   &otherwise.
\end{dcases*}
\end{equation} 

As illustrated in the Figure~\ref{fig:sequece}, for instance $T_{moving} =t_{5}-t_{4}$ is utilized to compute the relationship between the user's riding-off the elevator and turning-on the light. If $P(T_{moving})  >= \widetilde{p}$, THEN $E_{rideOffElevator}$ is conjunctive to $E_{lightOnRoom}$.

\section{Experiment}
\begin{figure}[h]
\centering
\includegraphics[height=1.3in, width=1\linewidth]{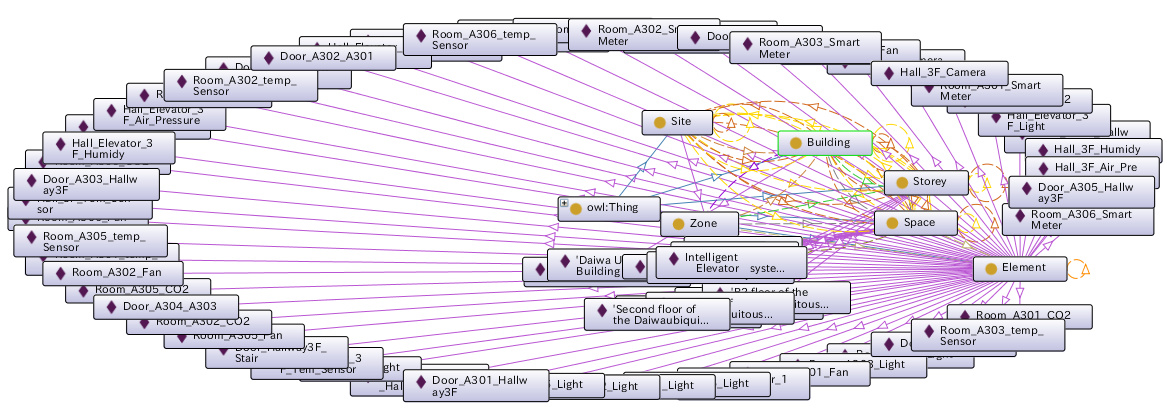}
\caption{The graph of entities and relationships shown in Protégé editor}
\label{fig:prete}
\end{figure}

We conducted experiments in a real smart building named ``Daiwa ubiquitous computing research building" at the university of Tokyo. The namespace of the ontology model is `\url{https://URI /daiwa_BOT\#}'. The building consists of 5 floors, including 3F, 2F, 1F, B1F and B2F,  with 43 space entities (room or hall, etc.). Sensors and actuators were defined as elements of the space and a total of 831 entries were modeled in the ontology. Figure~\ref{fig:prete} shows the visualization of the ontology structure.

We utilized the data collection of light switch and elevator movement records to infer knowledge about room occupancy and indoor users' trajectory. The experiment spanned 78-days, during which we collected 2667 light switch records and 14377 elevator movement records. Two types of use cases were simulated: 1. Timestamp-based detection of abnormal events related to light-off and light-on; 2. Time interval-based detection for events involving the conjunction of light switch and elevator usage.
\subsection{\textbf{Timestamp-based detection}}
\begin{table}[h]
\centering
 \caption{Probability statistics of the timestamp (in hour labeled in the left side) of light-off event}
\label{tab:lightoff}
\begin{tabular}{p{0.75cm}|p{0.35cm}|p{0.35cm}|p{0.35cm}|p{0.35cm}|p{0.35cm}|p{0.35cm}|p{0.35cm}|p{0.35cm}}
\toprule
Event  & \multicolumn{4}{c|}{Light-on} & \multicolumn{4}{c}{Light-off} \\\hline
Room    & 302 & 303  & 304  & 305   & 302  & 303  & 304  & 305  \\\hline
Median & 9.0 & 11.0 & 12.0 & 9.0  & 17.0 & 13.0 & 20.0 & 20.0\\
Std   & 2.17 & 3.00  & 3.83  & 3.94   & 1.16  & 4.34  & 6.07  & 7.77  \\\hline
0-1     & 0.00 & 0.00  & 0.00  & 0.00 & 0.00  & 0.00  & 0.00  & 0.01  \\
1-2   & 0.00 & 0.00  & 0.00  & 0.01  & 0.00  & 0.00  & 0.00  & 0.01  \\
2-3   & 0.00 & 0.00  & 0.00  & 0.01   & 0.00  & 0.00  & 0.00  & 0.01  \\
3-4    & 0.00 & 0.00  & 0.01  & 0.02   & 0.00  & 0.01  & 0.01  & 0.01  \\
4-5  & 0.01 & 0.01  & 0.01  & 0.03   & 0.00  & 0.01  & 0.01  & 0.02  \\
5-6  & 0.03 & 0.02  & 0.02  & 0.04  & 0.00  & 0.02  & 0.01  & 0.02  \\
6-7  & 0.07 & 0.03  & 0.03  & 0.05  & 0.00  & 0.02  & 0.01  & 0.02  \\
7-8  &\textbf{0.12} & 0.05  & 0.04  & 0.07  & 0.00  & 0.03  & 0.02  & 0.03  \\
8-9   & \textbf{0.16}& 0.08  & 0.06  & 0.08   & 0.00  & 0.04  & 0.02  & 0.03  \\
9-10   &\textbf{0.18}& \textbf{0.10}  & 0.07  & 0.09  & 0.00  & 0.06  & 0.03  & 0.03  \\
10-11   & \textbf{0.17} & \textbf{0.12}  & 0.09  & \textbf{0.10} & 0.00  & 0.07  & 0.03  & 0.04  \\
11-12  & \textbf{0.12} & \textbf{0.13}  & \textbf{0.10} & \textbf{0.10}   & 0.00  & 0.08  & 0.04  & 0.04  \\
12-13  & 0.07 & \textbf{0.13}  & \textbf{0.10}  & 0.09  & 0.00  & 0.09  & 0.05  & 0.05  \\
13-14  & 0.04 & \textbf{0.11}  & \textbf{0.10}  & 0.08  & 0.00  & 0.09  & 0.05  & 0.05  \\
14-15  & 0.01 & 0.08  & 0.09  & 0.07  & 0.02  & 0.09  & 0.06  & 0.05  \\
15-16  & 0.00 & 0.06  & 0.08  & 0.05  & \textbf{0.11}  & 0.09  & 0.06  & 0.05  \\
16-17  & 0.00 & 0.04  & 0.06  & 0.04  &\textbf{0.27} & 0.08  & 0.07  & 0.05  \\
17-18  & 0.00 & 0.02  & 0.05  & 0.03 & \textbf{0.33}& 0.06  & 0.07  & 0.05  \\
18-19  & 0.00 & 0.01  & 0.03  & 0.02  & \textbf{0.20}  & 0.05  & 0.07  & 0.05  \\
19-20 & 0.00 & 0.00  & 0.02  & 0.01  & 0.06  & 0.04  & 0.06  & 0.05  \\
20-21  & 0.00 & 0.00  & 0.01  & 0.01   & 0.01  & 0.03  & 0.06  & 0.05  \\
21-22   & 0.00 & 0.00  & 0.01  & 0.00   & 0.00  & 0.02  & 0.05  & 0.04  \\
22-23  & 0.00 & 0.00  & 0.00  & 0.00  & 0.00  & 0.01  & 0.05  & 0.04  \\
23-24 & 0.00 & 0.00  & 0.00  & 0.00  & 0.00  & 0.01  & 0.04  & 0.04 \\
\bottomrule
\end{tabular}
\end{table}

All the timestamps of light state switches in the rooms were extracted. The median and standard deviation values for several rooms, along with the probabilities for each one-hour time zone, are presented in Table~\ref{tab:lightoff}. Additionally, values exceeding 0.1 were emphasized. When comparing LightOff events, we observed that LightOn timestamps were more concentrated, showing a smaller standard deviation. In contrast, LightOff timestamps had larger standard deviation values, indicating greater variability. It is able to extract the semantics, for example, the timestamps of entering and exiting the student rooms of 304 and 305, were more scattered compared to other rooms.

\begin{figure}[h]
\centering
\includegraphics[height=2.7in, width=1\linewidth]{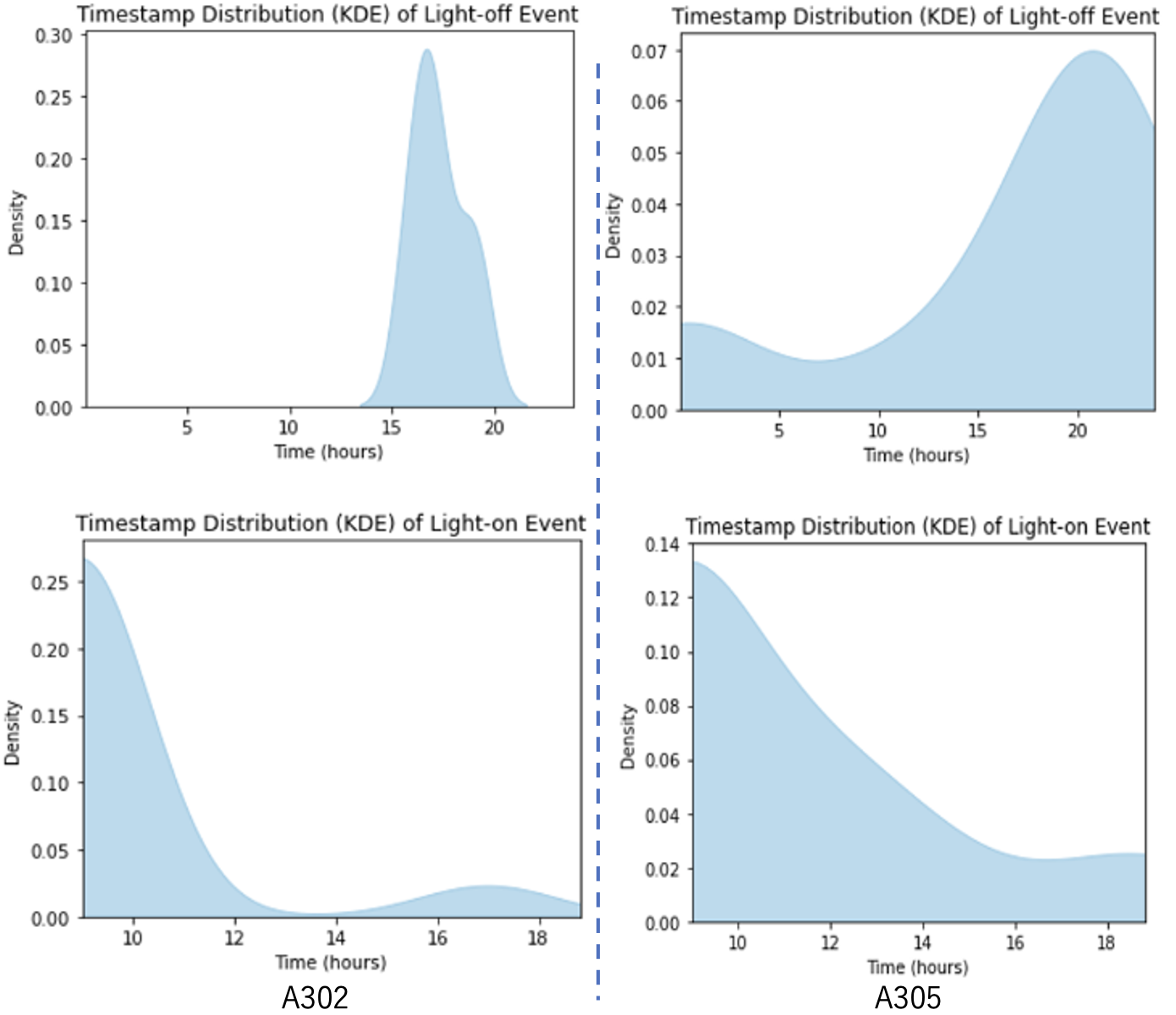}
\caption{Timestamp probability distribution of light-off event in A302 and A305}
\label{fig:2room-off}
\end{figure}

As an additional detail, two representative rooms were selected for room-usage comparison. As shown in the kernel density estimation in Figure \ref{fig:2room-off}, we observed significant differences in the timestamp distribution of LightOn and LightOff events. LightOff events in Room 302 are concentrated around 17:00, while those in Room 305 are concentrated around 20:00. This result aligns with the real usage scenario, given that the occupants of two rooms differ. Room 305 is predominantly occupied by students whereas Room 302 is mainly used by staff members. The entry and exit time records for students are relatively scattered, whereas those for staff members are relatively fixed due to regular work hours. The timestamp distributions of both LightOn and LightOff events align with the actual usage patterns of the rooms and the residents' habits of entering and exiting. 




\subsection{\textbf{Time interval-based detection}}
Based on timestamp-based event detection, the time interval between two corresponding events could be further approximated. For instance, a rule to detect the conjunction of lightOff and elevatorArriving can be formulated as follows: 

\textit{
If $Event_{lightOff}$ of $Room_{x}$ was detected at $t_{i}$; Then search the timestamp of $t_{j}$ denotes the time of $Event_{elevatorArriving}$ at $floor_{x}$ where $t_{j} > t_{i}$; 
Calculate $\Delta t = t_{j} - t_{i}$;
If $P(\Delta t)$ is $High$; Then the $Event_{LightOff}$ of $Room_{x}$ is Conjunctive to $Event_{elevatorArriving}$ ;
}

\begin{figure}[h]
\centering
\includegraphics[height=2.4in, width=\linewidth]{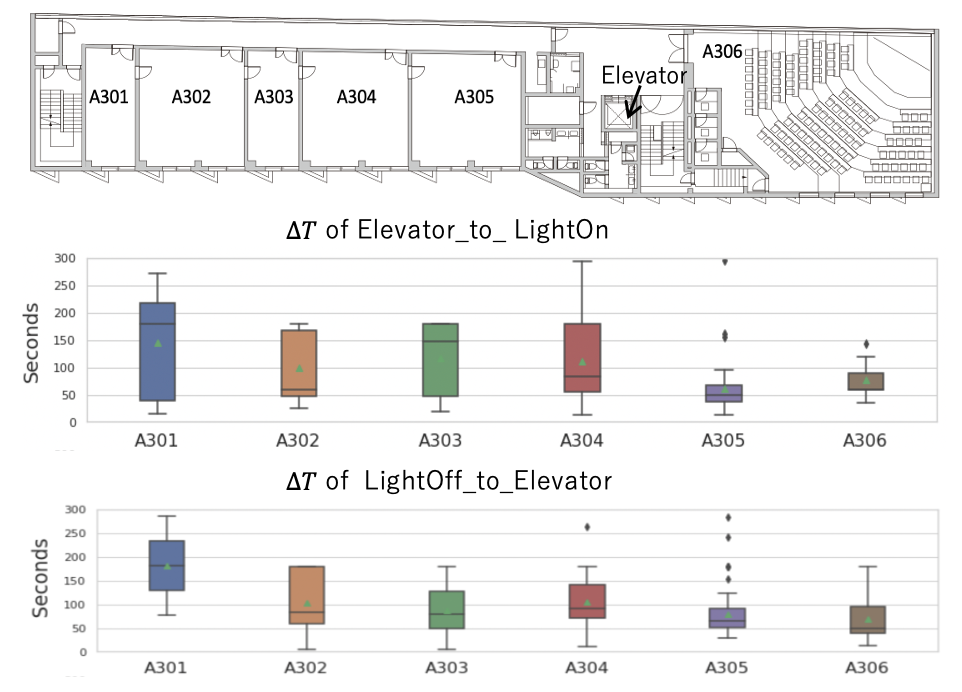}
\caption{Examples of approximated time interval $\Delta t$ of light-switch and elevator riding}
\label{fig:r2efloormap}
\end{figure}

As the first step for coarse filtering, we set $\Delta t <300$ to filter out noisy values of time intervals between ElevatorArriving and LightSwitch. Figure~\ref{fig:r2efloormap} shows time intervals between the event of several representative spaces with the floor map being showing. We observed the time intervals ($\Delta t$) are positively correlated with distances from the rooms to elevator. In particular, as indicated in the lower subgraph of Figure~\ref{fig:r2efloormap}, there is a stronger correlation in terms of time intervals and distance for the event conjunction of LightOff and ElevatorArriving.  

\subsubsection{Distance-based linear regression}
Based on the observation of time\_inteval and the corresponding distance. We propose a distance-based linear regression algorithm for optimization. As denoted in Equation~\ref{eq:dist}, $X_{1}$ is the count of hops between two space entities ($e_{i}$, $e_{j}$), representing the distance. These hops were extracted from the BOT graph using the terminology of 'bot:adjacentZone' and assigned as hops. $X_{2}$ represents numerical values for floors, assigned as numerical values of $[-2,-1,1,2,3]$. $Y$ represents the collected values of the time interval ($\Delta t$). The equation is linear, with $Y$ as the dependent variable and $X_{1}$ (hops of the ontology graph) and $X_{2}$ (floor values) as independent variable. The coefficients represent the weights of each independent variable in the linear combination. \begin{equation}
Y=\alpha+\beta_{1}\times X_{1}+\beta_{2}\times X_{2}\\ 
\label{eq:dist}
\end{equation}

Based on the distance values (hops) extracted from the BOT graph and the collected $\Delta t$ values, we implemented the linear regression model using the scikit-learn library \cite{pedregosa2011scikit}. Totally, 1093 time interval values were collected for training. The result of explicit intercept and coefficient values are listed in Table~\ref{tab:lrpa}, where $Y_{a}$ represents the time interval of LightOff and ElevatorArriving and $Y_{b}$ represents the one of ElevatorArriving and LightOn.

\begin{table}[h]
\centering
\begin{tabular}{l|lll}\toprule
  & $\alpha $& $\beta_{1}$ & $\beta_{2}$  \\\midrule
$Y_{a}$ ($LightOff \to ElevatorArriving$) & 49.36  & 12.86  &  2.94  \\
$Y_{b}$ ($ElevatorArriving \to LightOn $)& 29.81 & 14.74 &  7.29    \\
\bottomrule
\end{tabular}
\caption{The estimated results of the coefficients and intercept}
\label{tab:lrpa}
\end{table}



Based on the approximated values of $\Delta t$ and standard deviation ($\alpha $), we further calculate the value range to find the reasonable distribution of $\Delta t$. As the result presented in Table~\ref{tab:lr}, the time interval mean values ($\mu$) of event conjunction have been estimated. Meanwhile, the minimum and maximum values of time intervals were defined as :$\mu - K_{1}\times \alpha$ and  $\mu + K_{2}\times \alpha$, where $K_{1}$, $K_{2}$ were set to be 1 and 2, respectively. As an example, the time interval values of event conjunction of LightOff and ElevatorArriving has been presented in Table~\ref{tab:lr}, in which the room of which a total number of events over 80 were picked on the list. 
\begin{table}[h]
\centering
\caption{Time interval values optimized by linear regression model}
\begin{tabular}{cccccc}
\toprule
Space& $\Delta t$& Std ($\alpha $)&Min & Max\\\midrule
A305&81.16& 49.84 & 34.08&233.44\\
A304&95.90&84.36 &12.42&349.86\\
A202&118.10& 86.69&32.87&369.63\\
B202&88.96& 67.57&40.22&310.50\\
B204&44.73&48.02&21.18&213.26\\
\bottomrule
\end{tabular}
\label{tab:lr}
\end{table}

Using the optimized threshold range for fine-grained filtering, the event conjunction of LightOff and ElevatorArriving could be approximately inferred. As the result, the count of detected LightOff event of the room ($E_{r}$), event conjunction  of LightOff and ElevatorArriving (simplified as $E_{r}$->$E_{el}$) and their conditional probabilities have been summarized in Table~\ref{tab:evecount}.

\begin{table}[h]
\centering
\caption{Several inferred event-conjunction examples}
\begin{tabular}{ccccc}
\toprule
& \boldmath{$E_{r}$} & \boldmath{$E_{r}$}\textbf{->}\boldmath{$E_{el}$} &  \boldmath{$P_{r->el}$}\\
 \midrule
\textbf{Count} & \textbf{1331} & \textbf{471} & \textbf{0.35} \\\midrule
A305&110	&38	&0.345\\
A304&129 &32&0.248\\
A202&122& 22& \textbf{0.180}\\
B202&	90&	39&	0.433\\
B204&	89&	72&	\textbf{0.809}	\\
\bottomrule
\end{tabular}
\label{tab:evecount}
\end{table}

For the detail explanation, during the data collection period,  110 LightOff events have been detected in Room of A305, with 38 events detected in conjunction with the elevator riding-on, with the probability is 0.345. Meanwhile, the probability of Room A202 is 0.180, indicating that the occupants of A202 have a relatively low intention of taking a ride on the elevator. In contrast, the probability of Room B204 is 0.809, suggesting that the occupants in B204 prefer taking an elevator rather than using the stairs. The numerical result demonstrates distinct moving trajectories of users in different rooms. It also suggests potential applications such as automatic control on elevator. For instance, real-time monitoring of room occupancy could be achieved using the proposed timestamp-based event detection on the sensor observations. If the system detects a user who turned off the light with a high probability of using the elevator, it could trigger the elevator in advance for shortening users' waiting time.

\section{Conclusions}
In this paper, we presented a practical approach for time-probability dependent knowledge extraction method in IoT-enabled smart building. We proposed a unified API design as well as an ontology model for structuring the spatial features and dynamic status of IoT devices. Based on the ontology model, we proposed two types of time probability dependent knowledge extraction methods, one is timestamp-based method, the other is time interval-based method. 

The proposal has been implemented in a real smart building and 78-days data collection of the state on light and elevator has been conducted for evaluation. Timestamp-based light event detection and time interval-based event conjunctions on the light and elevator have been utilized for further inferring room occupancy and indoor users' moving trajectories. Informative numerical results shows the potentials for various application development for automatic control, such as elevator control for improving transport efficiency, etc. In terms of future work, long-term data collection  to ensure the stability of the probability distribution could be 
expected. 

\balance
\bibliography{IEEEabrv,ros}

\bibliographystyle{IEEEtran}


\end{document}